\documentclass[12pt,preprint]{aastex}
\usepackage{amsmath,euscript,amsfonts}

\newcommand{\beq}{\begin{equation}}
\newcommand{\eeq}{\end{equation}}
\newcommand{\ba}{\begin{array}}
\newcommand{\ea}{\end{array}}

\begin{document}


\lefthead{Liu et al.} \righthead{Origin of the different redshifts
among the broad hydrogen lines of quasars}

\title{
One possible explanation for the Balmer and Lyman line shifts in
quasars}

\author{\small \it Dangbo Liu\altaffilmark{1}, Jianrong Shi\altaffilmark{2},
 Junhan You\altaffilmark{1}
\affil{$^{1}$Center for Astronomy and Astrophysics, Department of
Physics and Astronomy, and Shanghai Key Lab for Particle Physics and
Cosmology, Shanghai Jiao Tong University, 800 Dongchuan Road,
 200240, Shanghai, China}
\affil{$^{2}$Key Laboratory of Optical Astronomy, National
Astronomical Observatories, Chinese Academy of Sciences, 100012,
Beijing, China}}
%

\begin{abstract}
%
Internal line shifts in quasars spectra have played a more prominent
role in our understanding of quasar structure and dynamics. The
observed different redshift among broad hydrogen lines is still an
amazing puzzle in the study of quasars. We have argued that the
broad hydrogen lines, as well as the low-ionization lines in
quasars, are significantly contributed by the Cerenkov quasi-line
emission of the fast electrons in the dense clouds/filaments/sheets
($N_{\rm H}\geq 10^{14}~{\rm cm^{-3}}$); whereas this line-like
radiation mechanism is invalid for producing the high ionization
lines. In order to account for redshift difference, the Cerenkov
line-like radiation mechanism could provide a plausible resolution:
it is the `Cerenkov line redshift', which is different from line to
line, causes the peculiar redshift-differences among Ly$\alpha$,
H$\alpha$ and H$\beta$ lines. The different redshifts among
different broad hydrogen lines could stand for an evidence to
quantitatively support that the observed broad hydrogen lines should
be blended by both the real line emission and the Cerenkov
quasi-line emission. The good fitting to the observed redshifts of
quasars confirms the existence of Cerenkov component in the broad
hydrogen lines, which indicates that, in the blended Ly$\alpha$
line, the line-intensity of the Cerenkov component approximately
equals that of the accompanying `normal line' (an approximate
equipartition of intensity between the two components in the broad
Ly$\alpha$ line). This result illustrates the importance of the
Cerenkov component in the broad lines of quasars, which can be
further confirmed by future observations.
\end{abstract}



\section{Introduction}
\label{sec:intro}

The emission lines are used to probe to the structure and dynamics
of the radiation region around the massive black holes in quasars.
In the early 1970s, peculiar Balmer line profiles had attracted
people's attentions, and all lines in a quasar do not produce the
same redshift, howbeit the redshift difference between emission
lines had not been systematically studied in quasars since the
1980s, even if some broad line shifts with respect to narrow lines
had been reported in quasars \citep{Osterbrock79}. The different
values for low-ionization lines (LILs) to be red-shifted relative to
high-ionization lines (HILs) has been well noted \citep{Gaskell82,
Wilkes86, Espey+89, Corbin90, Carswell+91, Sulentic+95,
McIntosh+99}. Recently, \citet{Vanden-Berk01} compiled a sample of
over 2200 quasars from the SDSS data and created a variety of
composite quasar spectra. The existence of slight redshifts of LILs
with respect to HILs is confirmed. The most peculiar result in their
composite spectra is suggesting a redshift difference between the
H$\alpha$ and H$\beta$ lines, $\Delta Z=Z_{\alpha}-Z_{\beta}\approx
10^{-4}$ \citep[see Table 4 of][]{Vanden-Berk01}. Observations show
this trend of $Z_{\alpha}>Z_{\beta}$ seems universal \citep[see
e.g., Table 4 in][also see, Tables 1 and 2 in Cheng et al. 1990 and
Table 2 in Nishihara et al. 1997]{Vanden-Berk01}. Over the past
decades, some physical and dynamical interpretations of line shifts
have been developed in the study of quasars such as atomic physics,
scattering processes, gravitational redshift, Doppler effect, etc.
\citep{Halenka15, Korista92, HE00, Ji12, DN79, Smith05, Laor06,
GG13, Netzer77, KG91, PW99, Shadmehri15}, however, just as described
in the review paper by \citet{Sulentic+16}, the systematical
redshift differences between emission lines in quasars have not been
fully clarified, and the structure and dynamics of the regions
emitting the low-ionization lines (LILs) are not fully understood.
The physical models associated dynamical processes in quasars need
to be further developed.

In our recent paper \citep{Liu+14}, we claimed that the Cerenkov
line-like radiation, created by the fast electrons in the dense
clouds/filaments/sheets with high densities $N_{\rm H}\geq
10^{14}~{\rm cm^{-3}}$, has a significant contribution to the broad
hydrogen lines and low-ionization lines of quasars. Therefore the
observed broad hydrogen line or low-ionization lines(LILs) should be
a `blended line' with two components: the `normal line', exactly at
$\lambda =\lambda_{lu}$, produced by the bound-bound transition
$u\rightarrow l$ in atoms/ions, and the `Cerenkov quasi-line'.
Actually, the latter is not a real emission line, instead, it is a
narrow continuum near the intrinsic wavelength $\lambda_{lu}$ of a
hydrogen line, $\lambda \gtrsim \lambda_{lu}$, hence the name
`Cerenkov line-like radiation', or simply , `Cerenkov quasi-line'.
Besides, we mentioned that the high ionization lines(HILs) (e.g.,
C~III, C~IV, N~V, etc.) are pure `normal lines', because the
Cerenkov radiation disappears in a fully ionized dense hydrogen
plasma, where all high-valence ions reside (e.g., C$^{++}$,
C$^{3+}$, N$^{4+}$, etc.). In this region, the Cerenkov mechanism
does not work due to the `effect of plasma oscillation', rendering
the refractive index of plasma $n<1$ \citep[see also,][]{Chen05}.
Therefore, supposing that the HILS are pure normal lines and the
LILs are blended lines with the two components of normal line and
Cerenkov quasi-line, this will be a plausible possibility to result
in the difference between LILs and HILs, represented by Mg~II
(2800~\AA) and C~IV (1549~\AA) respectively \citep{Gaskell82}
(detailed discussion in Section \ref{sec:Conclusion}). We should
indicate that the mechanism has been verified by laboratory
experiments \citep{Xu88, Yang89,Catravas+01}.

In this paper, we provide an observation evidence, quantitatively
supporting the above conclusions. We have mentioned that
\citep[see,][]{You84, You86, You00, Chen05}, one of the interesting
properties of the Cerenkov quasi-line is that, it does not exactly
locate at $\lambda =\lambda_{lu}$, but is slightly redshifted,
called as the `Cerenkov line redshift', which has a different amount
of redshift from line to line \citep{You84, You86, You00}. We try to
find evidences in observations to confirm the reality of this newly
recognized redshift. Our effort is successful. In Section
\ref{sec:model} of this paper, by using this new redshift effect and
based on the blend models, we calculate the redshifts of different
hydrogen lines to compare with the observations. The good fitting
very favors the proposition of the `blend line'.

Table \ref{tab01} of this paper (in Section \ref{sec:Calculation})
shows the averaged additional redshifts, $\Delta Z_{\rm
Ly\alpha}^{\rm obs}$, $\Delta Z_{\rm H\alpha}^{\rm obs}$, $\Delta
Z_{\rm H\beta}^{\rm obs}$, of $\sim 2,200$ quasars, relative to the
observed narrow line [O~III] 5007~{\AA} (see the first line in Table
\ref{tab01}, Section \ref{sec:Calculation}. Data are taken from
Table 4 in \citet{Vanden-Berk01}, where the redshift is in unit of
velocity. Note that, the redshift data of another strong hydrogen
line Ly$\beta$ is absent in their Table 4; hence absent in our Table
\ref{tab01}. But we still list the expected redshift of Ly$\beta$
line in the third line in Table \ref{tab01} for future detections in
observations). The observed redshifts of hydrogen lines Ly$\alpha$,
H$\alpha$ and H$\beta$ in the first line of Table \ref{tab01} are
really different from each other. As mentioned above, the Cerenkov
component in the proposed `blended line' has an additional `Cerenkov
line redshift' with different amount for different lines. It is
likely that the observed redshift-differences of different hydrogen
lines could arise from this additional redshift, rather than from a
velocity-origin, adopted in prevailing explanations.

Historically, it has been well known that, the low-ionization
lines(LILs) and the hydrogen lines of quasars have a systemic
redshift relative to the high-ionization lines(HILs)
\citep{Gaskell82, Wilkes86, Corbin90, Corbin91, Espey+89,
Carswell+91, Sulentic+95, McIntosh+99}. A large sample of over 2,200
quasars from the Sloan Digital Sky Survey (SDSS) database are
compiled and a variety of composite quasar spectra are obtained by
\citet{Vanden-Berk01}. They confirmed the existence of the slight
redshifs of LILs, as well as the slight blueshifts of HILs, with
respect to the narrow [O~III] 5007~{\AA} beyond all doubts.
According to the current explanation, both the redshifts of LILs and
the blueshifts of HILs are attributed to the stratified structure of
the broad line region (BLR) of quasars, with different radial
velocities of different ionization regions(inflow and outflow). From
the stratification models, it is inferred that there would be a
correlation between the velocity shifts and the ionization
potentials of spectral lines. This correlation is also confirmed by
using the same composite quasar spectra \citep[see Table 4
in][]{Vanden-Berk01}. However, an issue arises: why the
velocity-shifts are so different among the broad hydrogen lines, for
which the ionization potential is the same, $I_{\rm H}=13.6$ eV
\citep[see e.g., Table 4 in][also see, Tables 1 and 2 in Cheng et
al. 1990 and Table 2 in Nishihara et al. 1997]{Vanden-Berk01}. It
was suggested that, the hydrogen line stratification may occur due
to the different radiation transfer effects for different lines
\citep[e.g.,][]{OF06, Bentz2010}. The density gradient in BLR with
distance would cause the stratification of different hydrogen lines,
leading to the different velocity-shifts. This may be true, and
deserves to give a further detailed analysis to confirm this
viewpoint. But until now no quantitative fitting to the observed
redshift-differences of hydrogen lines has been developed by this
way. This situation promotes us to make an attempt to find
alternative solutions for this puzzle.

This paper is arranged as follows. In Section \ref{sec:model0}, we
describe the blend model for the broad lines of quasars. In Section
\ref{sec:Calculation}, we first give the theoretical redshifts of
the pure Cerenkov Ly$\alpha$, Ly$\beta$, H$\alpha$ and H$\beta$
lines, for preparing the subsequent calculation of the additional
redshifts of the blended lines. The detailed model calculations of
the redshifts of the blended hydrogen lines are presented in Section
\ref{sec:model}. Finally, conclusions and discussions are given in
Section \ref{sec:Conclusion}.

%


\section{Blend model for the broad lines --- Coexistence of two kinds of clouds/filaments with high and low densities in BLR}
\label{sec:model0}

We first give a clearer description for the blend model of the broad
lines of quasars, than in our recent paper \citep{Liu+14}. We have
mentioned that, the coexistence and blend of two kinds of emission
lines need a coexistence of two kinds of clouds/filaments with
higher and lower densities in the BLR. The clouds/filaments with
`standard' densities $N_{\rm H}\sim 10^{9}-10^{11}~{\rm cm}^{-3}$
\citep{DN79}, illuminated by the central UV and X-ray continuum, are
responsible to the normal line emission, but they have little
contribution to the Cerenkov lines(see Figure 1 in \citet{Liu+14}).
On the other hand, in the very dense clouds/filaments with $N_{\rm
H}\sim 10^{14}-10^{18}~{\rm cm}^{-3}$, the Cerenkov line-like
radiation occurs and becomes more prominent for higher densities;
whereas the normal lines totally disappear because of the
thermalization in the dense gas (when $N_{\rm H}> 10^{13}~{\rm
cm}^{-3}$, see \citet{RNF89}). In brief, the normal lines and the
Cerenkov quasi-lines, respectively, arise from the `standard' and
the dense clouds/filaments in the BLR (intermediate densities of
$10^{11}~{\rm cm}^{-3}< N_{\rm H}< 10^{14}~{\rm cm}^{-3}$ are also
possible---it is unfair to exclude their existence,
though they are unhelpful for producing the line-emission).

The `standard' and the dense clouds/filaments, coexisting in BLR,
are likely to distribute in different regions. It is a consensus
that the `standard' clouds/filaments spread in the whole BLR, while
a vast quantity of very dense cloudlets or filaments/sheets with
small sizes are scattered in the very inner portion of BLR, most
plausibly in the magnetosphere around the central engine---typically
on scales of $d\lesssim 100~R_{g}$ ($R_{g}=GM/c^{2}$ is the
gravitational radius of a central mass $M$), where the random
magnetic fields of $10^{3}-10^{4}$ Gauss may pervade \citep{Rees87,
FR88, GR88, LW88, Celotti92, ST93, Collin96, Kuncic96, Kuncic97,
CD98, CR99, Lawrence12}. The magnetic fields are likely to have
complex structures. The dense cloudlets/filaments in the
magnetosphere can be supported and confined by the magnetic fields.
These dense cloudlets, with temperature of $\sim 10^{4}-10^{5}$~K,
reprocess the primary non-thermal radiation of relativistic
electrons from the central engine, and then  contribute a
quasi-blackbody continuum in the optical-UV bands (the `big blue
bump'). A possible range of the density $N_{\rm H}$ and the size
scale $r$ for dense cloudlets/filaments/sheets were suggested by the
above authors; $N_{\rm H}\sim 10^{14}-10^{18}~{\rm cm}^{-3}$, $r\sim
10^{2}-10^{9}~{\rm cm}$ (for filaments and sheets, $r$ expresses the
thickness). \citet{Kuncic96} further suggested that the most
suitable values, favoring to surviving of the dense clouds around
the central engine, are $N_{\rm H}\sim 10^{16}-10^{18}~{\rm
cm}^{-3}$ and $r\sim 10^{2}-10^{6}~{\rm cm}$. Besides, the presence
of abundant fast electrons in the innermost region is conceivable.
All these conditions are favorable to the production of the
efficient Cerenkov line-like radiation. In brief, in our blend
models, the dense cloudlets and the `standard' cloudlets,
respectively, exist in different regions of the BLR; the former are
scattered in the inner magnetosphere around the central engine,
whereas the latter distribute in the whole region of BLR.
Some observations support the above picture of different regions:
\citet{Bentz2010} reported that, the variation response of hydrogen
lines to the UV-continuum in Arp~151 shows a deficit of prompt
response in the Balmer-line cores but strong prompt response in the
red wings of lines. This is easy to understand if the cores and the
red-wings of lines are, respectively, dominated by the normal line
and the Cerenkov line emissions; and if the latter is produced by
the dense matter in the innermost region of BLR.

Moreover, we mention that, owing to the effect of the combined
forces of gravity, radiation pressure and magnetic stresses, the
dense clouds/filaments/sheets inside the magnetosphere should
markedly deviate from the virial motion, controlled by the gravity
alone, thus the speeds of clouds/filaments should be much less than
the virial velocity due to the damping of both the radiation and the
magnetic viscosity, i.e., $v\ll v_{\rm virial}\approx v_{\rm
ff}=c\left (r/r_{\rm g}\right )^{-1/2}$, $r_{\rm g}=GM/c^{2}$ is the
gravitational radius with a central mass $M$. The possibility of
$v<v_{\rm virial}$ is confirmed by the observations: the observed
broadening of the hydrogen lines, emitted simultaneously with the
central flare, is markedly less than that given by the virial
theorem (see the observed line widths in regions of time-delay
$\tau\approx 0$ in Figures 3 and 4 in \citet{Bentz2010}). In the
following calculations, we neglect both the velocity-broadening and
velocity-shift of the Cerenkov lines, caused by the un-virial, slow
flailing of dense clouds/filaments in the central magnetosphere. We
assume that, all of the additional redshifts of hydrogen lines arise
from the `Cerenkov line redshift'.



\section{Redshifts of the pure Cerenkov Ly$\alpha$, Ly$\beta$,
H$\alpha$, H$\beta$ lines} \label{sec:Calculation}

Before the model calculations for the blended hydrogen lines, we
first present  the additional slight redshifts of pure Cerenkov
hydrogen lines. For a specific Cerenkov quasi-line near the
intrinsic wavelength $\lambda\gtrsim\lambda_{lu}$, the `Cerenkov
line redshift' is given by \citep{You84, You86, You00}
\begin{eqnarray}
\Delta Z_{\lambda_{lu}}^{\rm Cer}\equiv \frac{\Delta\lambda_{p}^{\rm Cer}}{\lambda_{lu}}
=1.04\times 10^{-11}\sqrt{\lambda_{lu}A_{ul}\Gamma_{lu}g_{u}\left (\frac{N_{l}}{g_{l}}-\frac{N_{u}}{g_{u}}\right )N_{p}^{-1}p^{5}}
~~~~({\rm valid~for}~N_{\rm H}>10^{15}~{\rm cm}^{-3}).
\label{Cer-redshift}
\end{eqnarray}
Here, $\lambda_{lu}$ is the intrinsic wavelength of a specific
hydrogen line with the upper level $u$ and lower level $l$; $A_{ul}$
is the Einstein's spontaneous emission coefficient for the
transition $u\rightarrow l$; $\Gamma_{lu}\equiv
\Gamma_{l}+\Gamma_{u}=\sum\limits_{i<l}A_{li}+\sum\limits_{j<u}A_{uj}$
is the total quantum damping constant of the specific hydrogen line
with wavelength $\lambda_{lu}$, which is related to Einstein's
spontaneous emission probabilities $A_{li}$ and $A_{uj}$. $g_{u}$
and $g_{l}$ are the degeneracy of the upper level $u$ and the lower
level $l$, respectively. $N_{u}$ and $N_{l}$ represent,
respectively, the number densities of neutral hydrogen at the upper
and the lower levels. $N_{p}$ is the number density of neutral
hydrogen at the lowest level $p$ of the photoelectric absorption,
which gives the dominant absorption among all qualified
photoelectric levels of a hydrogen atom for the incident line photon
with wavelength $\lambda_{lu}$. For the Cerenkov Ly$\alpha$ and
Ly$\beta$ lines, $p=2$; for the Cerenkov H$\alpha$ and H$\beta$
lines, $p=3$, etc. $\lambda_{lu}$ is in unit cm.

From equation (\ref{Cer-redshift}), we obtain the additional
redshifts of the Cerenkov Ly$\alpha$, Ly$\beta$, H$\alpha$ and
H$\beta$ lines:
\begin{eqnarray}
\Delta Z_{\rm Ly\alpha}^{\rm Cer}& = & 1.93\times 10^{-4}\left ( \frac{N_{1}}{N_{2}}-\frac{1}{4}\right )^{1/2},\nonumber \\
\Delta Z_{\rm Ly\beta}^{\rm Cer}& = & 4.22\times 10^{-5}\left ( \frac{N_{1}}{N_{2}}-\frac{1}{9}\frac{N_{3}}{N_{2}}\right )^{1/2},\nonumber \\
\Delta Z_{\rm H\alpha}^{\rm Cer} & = & 3.13\times 10^{-4}\left (\frac{N_{2}}{N_{3}}-\frac{4}{9}\right )^{1/2} , \nonumber \\
\Delta Z_{\rm H\beta}^{\rm Cer} & = & 1.47\times 10^{-4}\left (\frac{N_{2}}{N_{3}}-\frac{1}{4}\frac{N_{4}}{N_{3}}\right )^{1/2},
\label{Cer-redshift1}
\end{eqnarray}
where $N_{1}$, $N_{2}$, $N_{3}$ and $N_{4}$ represent the number
densities of neutral hydrogen in levels $n=1,~2,~3$ and $4$,
respectively. Here we adopt atomic parameters in Equation
(\ref{Cer-redshift}) as $\lambda_{12}=1.216\times 10^{-5}$~cm,
$A_{21}=4.70\times 10^{8}~{\rm s}^{-1}$, $\Gamma_{12}=A_{21}$,
$\lambda_{13}=1.026\times 10^{-5}$~cm, $A_{31}=5.57\times
10^{7}~{\rm s}^{-1}$, $\Gamma_{13}=9.98\times 10^{7}~{\rm s}^{-1}$,
$\lambda_{23}=6.563\times 10^{-5}$~cm, $A_{32}=4.41\times
10^{7}~{\rm s}^{-1}$, $\Gamma_{23}=5.70\times 10^{8}~{\rm s}^{-1}$,
$\lambda_{24}=4.861\times 10^{-5}$~cm, $A_{42}=8.42\times
10^{6}~{\rm s}^{-1}$, $\Gamma_{24}=5.00\times 10^{8}~{\rm s}^{-1}$,
$g_{1}=2$, $g_{2}=8$, $g_{3}=18$, and $g_{4}=32$ for our
calculations.

We have mentioned that \citep{Liu+14}, the Cerenkov line-like
radiation is effective only for the dense cloudlets (and/or
filaments, sheets) with high densities $N_{\rm H}\geq 10^{14}~{\rm
cm}^{-3}$. For the dense clouds, the populations of hydrogen,
$N_{1}$, $N_{2}$, $N_{3}$ and $N_{4}$, contained in Equation
(\ref{Cer-redshift1}), are given by the Boltzmann law, thus
$N_{1}/N_{2}=\left (g_{1}/g_{2}\right )\exp\left (10.2~{\rm
eV}/kT\right )$, $N_{2}/N_{3}=\left (g_{2}/g_{3}\right )\exp\left
(1.88~{\rm eV}/kT\right )$, and $N_{4}/N_{3}=\left
(g_{4}/g_{3}\right )\exp\left (-0.66~{\rm eV}/kT\right )$. Therefore
the temperature $T$ of the dense cloud/filament/sheet becomes an
unique parameter for determining the ratios $N_{1}/N_{2}$,
$N_{2}/N_{3}$ and $N_{4}/N_{3}$, hence for the redshifts $\Delta
Z_{\rm Ly\alpha}^{\rm Cer}$, $\Delta Z_{\rm Ly\beta}^{\rm Cer}$,
$\Delta Z_{\rm H\alpha}^{\rm Cer}$ and $\Delta Z_{\rm H\beta}^{\rm
Cer}$ in Equation (\ref{Cer-redshift1}).

We obtain the permitted range of $T$ of the dense matter in quasars
as follows: As mentioned in Section \ref{sec:model0}, a vast amount
of very dense and `cold' clouds with typical temperature $T$ of
$\gtrsim 10^{4}$~K distribute in the magnetosphere around the
central engine, which reprocess the non-thermal radiation from the
central engine and contribute the `ultraviolet excess' ( the `big
blue bump' ) in the optical-ultraviolet continuum of quasars, via
the optically thick thermal emission. This thermal component can be
well described by a blackbody at a single temperature $T$.
\citet{MS82} successively fitted the observed spectra of ultraviolet
excess of eight Seyfert 1 and quasars, and obtained the blackbody
temperatures $T$ in range $2\times 10^{4}~{\rm K}\lesssim T\lesssim
3\times 10^{4}~{\rm K}$ (see Table 2 in \citet{MS82}). The median or
average temperature is $T\approx 2.5\times 10^{4}$~K. They stressed
that the thermal component need not be a single temperature
blackbody. This is only the simplest form to fit the data well. In
reality, there could be small contributions present from hotter or
cooler gas which would be difficult to detect. Furthermore, the
observation indicates that there is little thermal gas much hotter
than 40,000~K \citep{Green+80}. Therefore, in the following, we
cease the calculations at $T=4\times 10^{4}$~K, and take $2\times
10^{4}~{\rm K}\lesssim T\lesssim 3\times 10^{4}~{\rm K}$ as a
permitted range of the temperature $T$.

\begin{figure}[ht]
\centerline{\includegraphics[width=12cm]{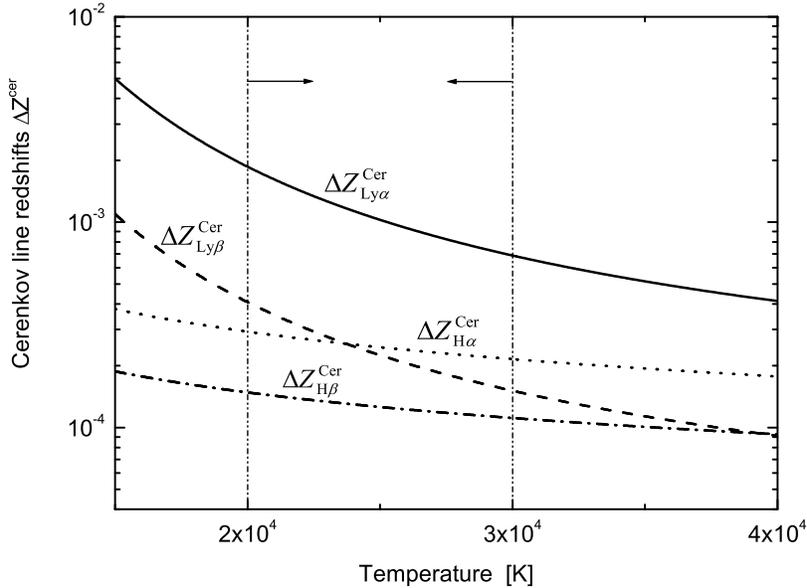}}
\caption{The calculated redshifts of Cerenkov Ly$\alpha$, Ly$\beta$,
H$\alpha$ and H$\beta$ lines under different temperature of the dense
clouds in quasars, shown by the solid, dashed, dotted and dash-dotted lines, respectively.
The vertical dash-dot-dotted lines label the permitted
temperature range of dense matter, $2\times 10^{4}~{\rm K}\lesssim T\lesssim 3\times
10^{4}~{\rm K}$, which is obtained
by fitting the `big blue bump' in the observed optical-UV continua of quasars \citep{MS82}
.
}
\nonumber
\label{fig:1}
\end{figure}

By using equation (\ref{Cer-redshift1}), the calculated curves of
Cerenkov line redshifts $\Delta Z_{\rm Ly\alpha}^{\rm Cer}\sim T$,
$\Delta Z_{\rm Ly\beta}^{\rm Cer}\sim T$, $\Delta Z_{\rm
H\alpha}^{\rm Cer}\sim T$ and $\Delta Z_{\rm H\beta}^{\rm Cer}\sim
T$, are shown in Figure \ref{fig:1} by the solid, dashed, dotted and
dash-dotted lines respectively, taking the temperature $T$ of the
dense clouds/filaments/sheets in quasars as a free variable. In
Figure \ref{fig:1}, the permitted range of $T$ is labeled by the
vertical dash-dot-dotted lines. In this region, the calculated
Cerenkov redshifts are restricted in ranges $6.88\times
10^{-4}\lesssim \Delta Z_{\rm Ly\alpha}^{\rm Cer}\lesssim 1.86\times
10^{-3}$, $1.51\times 10^{-4}\lesssim \Delta Z_{\rm Ly\beta}^{\rm
Cer}\lesssim 4.07\times 10^{-4}$, $2.15\times 10^{-4}\lesssim \Delta
Z_{\rm H\alpha}^{\rm Cer}\lesssim 2.93\times 10^{-4}$, and
$1.11\times 10^{-4}\lesssim \Delta Z_{\rm H\beta}^{\rm Cer}\lesssim
1.48\times 10^{-4}$, respectively.

We mention that, actually, the observed $\Delta Z_{\rm
Ly\alpha}^{\rm obs}$, $\Delta Z_{\rm H\alpha}^{\rm obs}$ and $\Delta
Z_{\rm H\beta}^{\rm obs}$, listed in Table \ref{tab01}, are the
averaged values of $\sim 2,200$ quasars \citep{Vanden-Berk01}.
Therefore it is more reasonable to make a theoretical calculation of
redshifts at the average temperature $T$, and then to compare with
the average observed ones. Taking the median temperature $T\approx
2.5\times 10^{4}$~K ( $kT\approx 2.2$~eV) in range $2\times
10^{4}~{\rm K}\lesssim T\lesssim 3\times 10^{4}~{\rm K}$, the
corresponding median values of redshifts $\Delta Z_{\rm
Ly\alpha}^{\rm Cer}$, $\Delta Z_{\rm Ly\beta}^{\rm Cer}$, $\Delta
Z_{\rm H\alpha}^{\rm Cer}$ and $\Delta Z_{\rm H\beta}^{\rm Cer}$ are
listed in the second line in Table \ref{tab01}.

\begin{table}[htbp]
\caption{Observed and theoretically calculated values of the
additional slight redshifts of Ly$\alpha$, Ly$\beta$, H$\alpha$,
H$\beta$ lines relative to [O~III] 5007~{\AA} }
\begin{tabular}
{ccccc} \hline\hline \raisebox{-1.50ex}[0cm][0cm]{}&
 ~~Ly$\alpha$~~  & ~~Ly$\beta$~~ & ~~H$\alpha$~~ & ~~H$\beta$~~ \\
\hline \raisebox{-0.50ex}[0.3cm][0.4cm] {Observed mean}& $\Delta Z_{\rm Ly\alpha}^{\rm obs}(10^{-4})$ & $\Delta Z_{\rm Ly\beta}^{\rm obs}(10^{-4})$ & $\Delta Z_{\rm H\alpha}^{\rm obs}(10^{-4})$
& $\Delta Z_{\rm H\beta}^{\rm obs}(10^{-4})$
\\\raisebox{-0.50ex}[0cm][0cm] {redshift values\footnote{}}
& $(4.77\pm 3.03) $ & --- &  $(0.93\pm 0.43)$ & $(0\pm 0.50)$ \\
\hline \raisebox{-0.50ex}[0.3cm][0.4cm] {Redshifts of Cerenkov}& $\Delta Z_{\rm Ly\alpha}^{\rm Cer}(10^{-4})$ & $\Delta Z_{\rm Ly\beta}^{\rm Cer}(10^{-4})$ & $\Delta Z_{\rm H\alpha}^{\rm Cer}(10^{-4})$
& $\Delta Z_{\rm H\beta}^{\rm Cer}(10^{-4})$
\\\raisebox{-0.50ex}[0cm][0cm] {hydrogen lines\footnote{}}
& $9.77$ & $2.14$ & $2.42$ & $1.24$ \\
\hline \raisebox{-0.50ex}[0.3cm][0.4cm] {Expected redshifts of} & $\Delta Z_{\rm Ly\alpha}^{\rm Exp}(10^{-4})$ & $\Delta Z_{\rm Ly\alpha}^{\rm Exp}(10^{-4})$ & $\Delta Z_{\rm H\alpha}^{\rm Exp}(10^{-4})$
& $\Delta Z_{\rm H\beta}^{\rm Exp}(10^{-4})$
\\\raisebox{-0.50ex}[0cm][0cm] {the blended hydrogen}
& $4.88$ & $1.07$ & $1.21$ & $0.62$ \\
\raisebox{-0.50ex}[0cm][0cm] {lines\footnote{} (blend model 1)}
& &  &  &  \\
\hline \raisebox{-0.50ex}[0.3cm][0.4cm] {Expected redshifts of} & $\Delta Z_{\rm Ly\alpha}^{\rm Exp}(10^{-4})$ & $\Delta Z_{\rm Ly\alpha}^{\rm Exp}(10^{-4})$ & $\Delta Z_{\rm H\alpha}^{\rm Exp}(10^{-4})$
& $\Delta Z_{\rm H\beta}^{\rm Exp}(10^{-4})$
\\\raisebox{-0.50ex}[0cm][0cm] {the blended hydrogen}
& $4.76$ & --- & $0.94$ & $0$ \\
\raisebox{-0.50ex}[0cm][0cm] {lines\footnote{} (blend model 2)}
& &  &  &  \\
\hline
\label{tab01}
\end{tabular}
~~~~{\small Notes. --- 1.~The observation data are taken from Table
4 in \citet{Vanden-Berk01} with a large data set of over 2200
quasars. 2.~In this calculation, we take a median temperature
$T\approx 2.5\times 10^{4}$~K (or $kT\approx 2.2$~eV) in the
permitted temperature range of dense clouds, $2\times 10^{4}~{\rm
K}\lesssim T\lesssim 3\times 10^{4}~{\rm K}$\citep[for calculation
details see Section \ref{sec:Calculation}]{MS82}. 3.~For the blend
model 1, we adopt a simple equipartition of intensity between the
normal line and the Cerenkov line, $I_{\lambda_{lu}}^{\rm
nor}\approx I_{\lambda_{lu}}^{\rm Cer}$ (for details, see Section
\ref{sec:model}). 4.~For the blended model 2, we take the
intensity-ratio $I_{\rm Ly\alpha}^{\rm Cer}/I_{\rm Ly\alpha}^{\rm
nor}=0.95$, $I_{\rm H\alpha}^{\rm Cer}/I_{\rm H\alpha}^{\rm
nor}=0.63$ and $I_{\rm H\beta}^{\rm Cer}/I_{\rm H\beta}^{\rm
nor}=0$, to get the best fitting with the observed redshifts
values.}
\end{table}

We notice that, the above calculated redshifts of the pure Cerenkov
lines, listed in the second line of Table \ref{tab01}, are already
comparable with the observed values, with the same orders of
magnitude of $\Delta Z\approx 10^{-4}$ and same un-equality sequence
$\Delta Z_{\rm Ly\alpha}>\Delta Z_{\rm H\alpha} >\Delta Z_{\rm
H\beta}$. This is unlikely a coincidence, though the discrepancy
between the theoretical and the observational values can not be
ignored.


\section{Model calculation of redshifts of the blended hydrogen lines}
\label{sec:model}

Table \ref{tab01} shows a marked deviation of the redshifts of pure
Cerenkov hydrogen Ly$\alpha$, H$\alpha$ and H$\beta$ lines from the
observed values, though with the same order of magnitude $\Delta
Z\sim 10^{-4}$. The theoretical redshifts are higher than the
corresponding observed ones, $\Delta Z_{\rm Ly\alpha}^{\rm Cer} >
\Delta Z_{\rm Ly\alpha}^{\rm obs}$, $\Delta Z_{\rm H\alpha}^{\rm
Cer} > \Delta Z_{\rm H\alpha}^{\rm obs}$ and $\Delta Z_{\rm
H\beta}^{\rm Cer} > \Delta Z_{\rm H\beta}^{\rm obs}$. This is easy
to understand in the frame of the `blend models'. As a blended line,
the observed broad hydrogen line contains another important
component--the `normal line', which has not been taken into account
yet (note that, in the composite spectra of \citet{Vanden-Berk01},
both the broad and the narrow `normal lines' (the bound-bound
transition lines) are all involved in the observed broad hydrogen
lines). Differing to the Cerenkov component, the `normal line'
exactly locates at $\lambda =\lambda_{lu}$, with no additional
redshift ($\Delta Z_{\lambda_{lu}}^{\rm nor}=0$), which inevitably
leads to a marked decrease of the resultant redshift of the `blended
line'.

The expected redshift of a blended line is determined by the
fraction of the Cerenkov component in the total line intensity,
\begin{eqnarray}
\Delta Z_{\lambda_{lu}}^{\rm exp}&=&\left (\frac{1}{I_{\lambda_{lu}}^{\rm nor}
+I_{\lambda_{lu}}^{\rm Cer}} \right ) \left (I_{\lambda_{lu}}^{\rm nor}\Delta Z_{\lambda_{lu}}^{\rm nor}
+I_{\lambda_{lu}}^{\rm Cer}\Delta Z_{\lambda_{lu}}^{\rm Cer}\right ) \nonumber \\
&=&\left (\frac{I_{\lambda_{lu}}^{\rm Cer}}{I_{\lambda_{lu}}^{\rm nor}+I_{\lambda_{lu}}^{\rm Cer}} \right )
\Delta Z_{\lambda_{lu}}^{\rm Cer}\qquad ({\rm where}~ \Delta Z_{\lambda_{lu}}^{\rm nor}=0),
\label{expectedredshift0}
\nonumber
\end{eqnarray}
or
\begin{eqnarray}
\Delta Z_{\lambda_{lu}}^{\rm exp}=\left (\frac{I_{\lambda_{lu}}^{\rm Cer}/I_{\lambda_{lu}}^{\rm nor}}
{1+I_{\lambda_{lu}}^{\rm Cer}/I_{\lambda_{lu}}^{\rm nor}} \right )
\Delta Z_{\lambda_{lu}}^{\rm Cer}~.
\label{expectedredshift}
\end{eqnarray}
where $I_{\lambda_{lu}}^{\rm Cer}$ and $I_{\lambda_{lu}}^{\rm nor}$
are the line-intensities of the Cerenkov quasi-line and the normal
line, respectively. The total intensity of the blend line is
$I_{\lambda_{lu}}^{\rm total}=I_{\lambda_{lu}}^{\rm
nor}+I_{\lambda_{lu}}^{\rm Cer}$. Equation (\ref{expectedredshift})
is a good approximation for describing the resultant redshift of a
blended line, where the peak of the Cerenkov line quasi-line is very
closed to the accompanying normal line.

In order to obtain the expected redshift of a blended line, it is
necessary to know the intensity-ratio $I_{\lambda_{lu}}^{\rm
Cer}/I_{\lambda_{lu}}^{\rm nor}$ of the two components. This could
be a tedious work, depending on a detailed data-analysis for both
the observed intensities and the redshifts of different hydrogen
lines of quasars. We plan to do it in near future. In this paper, we
tentatively take the intensity-ratio in Equation
(\ref{expectedredshift}) as a modulated parameter to calculate
$\Delta Z_{\lambda_{lu}}^{\rm exp}$, and then to compare with
observations. We envisage the values of intensity-ratio
$I_{\lambda_{lu}}^{\rm Cer}/I_{\lambda_{lu}}^{\rm nor}$ in two
manners, which we call as the `blend model 1' and the `blend model
2' in the following, respectively.

\subsection{Blend model 1}
\label{sec:blend-model1}

As a primary approximation, we simply assume that, the average
intensity-ratio in the blend line of quasars is near to unit,
\begin{eqnarray}
I_{\lambda_{lu}}^{\rm Cer}/I_{\lambda_{lu}}^{\rm nor}\approx 1~,
\label{intensityratio}
\end{eqnarray}
i.e., we have an `equipartition of intensity' between the two
components in the blend lines. Although so far we can not give a
convincing argument to show the reasonableness of the approximate
Equation (\ref{intensityratio}), but we have emphasized the
importance of the Cerenkov component in quasars. We mentioned that
the Cerenkov quasi-line is strong enough to compete with the
accompanying normal line in the blended hydrogen lines or
low-ionization lines \citep{Liu+14}. This implies that the
intensity-ratio of the Cerenkov quasi-line to the normal line should
be not far from unit, e.g., in a narrower range $0.5<
I_{\lambda_{lu}}^{\rm Cer}/I_{\lambda_{lu}}^{\rm nor} < 1.5$.
Therefore, in the primary $\Delta Z_{\lambda_{lu}}^{\rm
exp}-$calculations, the approximate Equation \ref{intensityratio},
$I_{\lambda_{lu}}^{\rm Cer}\approx I_{\lambda_{lu}}^{\rm nor}$,
seems to be acceptable . Inserting Equation (\ref{intensityratio})
into Equation (\ref{expectedredshift}), we obtain
\begin{equation}
\Delta Z_{\lambda_{lu}}^{\rm exp}\approx \frac{1}{2}\Delta Z_{\lambda_{lu}}^{\rm Cer}
\qquad ({\rm when}~I_{\lambda_{lu}}^{\rm Cer}/I_{\lambda_{lu}}^{\rm nor}\approx 1)
\label{expectedredshift1}
\end{equation}

By using Equation (\ref{expectedredshift1}), the expected resultant
redshifts of the blended hydrogen lines Ly$\alpha$, Ly$\beta$,
H$\alpha$ and H$\beta$ are shown in the third line in Table
\ref{tab01}. Obviously, the expected redshifts, given by the `blend
model 1', are much closer to the observed ones (Table \ref{tab01},
the first line), than the pure Cerenkov lines (the second line in
Table \ref{tab01}). This result strongly supports the reasonableness
of our proposition of the blended hydrogen lines.

However, there still exist remarkable discrepancies between the
predictions of model 1 and the observations for the H$\alpha$ and
H$\beta$ lines (particularly for H$\beta$), indicating that the
approximation of `equipartition of intensity' (Equations
(\ref{intensityratio}) and (\ref{expectedredshift1})) is not equally
good for all hydrogen lines. This is easy to understand. In fact,
the Cerenkov and the normal line emissions are in principle
different mechanisms, independent to each other. Generally, the
series of intensity-ratios of the Cerenkov hydrogen lines, $I_{\rm
Ly\alpha}^{\rm Cer}/I_{\rm Ly\beta}^{\rm Cer}/I_{\rm H\alpha}^{\rm
Cer}/I_{\rm H\beta}^{\rm Cer}$ should be different to $I_{\rm
Ly\alpha}^{\rm nor}/I_{\rm Ly\beta}^{\rm nor}/I_{\rm H\alpha}^{\rm
nor}/I_{\rm H\beta}^{\rm nor}$ for the normal lines. Therefore,
$I_{\rm Ly\beta}^{\rm Cer}\neq I_{\rm Ly\beta}^{\rm nor}$, $I_{\rm
H\alpha}^{\rm Cer}\neq I_{\rm H\alpha}^{\rm nor}$, and $I_{\rm
H\beta}^{\rm Cer}\neq I_{\rm H\beta}^{\rm nor}$, etc. even if
$I_{\rm Ly\alpha}^{\rm Cer} = I_{\rm Ly\alpha}^{\rm nor}$ is already
given.

\subsection{Blend model 2}
\label{sec:blend-model2}

In this model, we abandon Equations (\ref{intensityratio}) and
(\ref{expectedredshift1}), tentatively take the line-intensity ratio
$I_{\lambda_{lu}}^{\rm Cer}/I_{\lambda_{lu}}^{\rm nor}$ in Equation
(\ref{expectedredshift}) as a modulated parameter to get the best
fitting to each observed redshifts for Ly$\alpha$, H$\alpha$ and
H$\beta$ lines. Taking $I_{\rm Ly\alpha}^{\rm Cer}/I_{\rm
Ly\alpha}^{\rm nor}=0.95$, $I_{\rm H\alpha}^{\rm Cer}/I_{\rm
H\alpha}^{\rm nor}=0.63$, $I_{\rm H\beta}^{\rm Cer}/I_{\rm
H\beta}^{\rm nor}=0$, from Equation (\ref{expectedredshift}), we
obtain the best expected redshifts of Ly$\alpha$, H$\alpha$ and
H$\beta$ lines, shown in the fourth line in Table 1, which are in
good consistence with the observed values. The good fitting shows
the plausibility of above adopted intensity-ratios in quasars,
though the theoretical argument has not been given yet.

But we would like to mention that, the chosen value $I_{\rm
H\beta}^{\rm Cer}/I_{\rm H\beta}^{\rm nor}=0$ for fitting the `zero
redshift' of H$\beta$ line seems to be quite questionable. The
observed $\Delta Z_{\rm H\beta}^{\rm obs} = (0\pm 0.50)\times
10^{-4}$, with the largest error of measurement among Ly$\alpha$,
H$\alpha$ and H$\beta$ lines, only indicates that no additional
redshift can be detected at the level of order of magnitude $\Delta
Z\sim 10^{-4}$; and the observed redshift is uncertain in a wide
range, $0\leq \Delta Z_{\rm H\beta}^{\rm obs}\leq 0.5\times
10^{-4}$, rather than $\Delta Z_{\rm H\beta}^{\rm obs}=0$.
Correspondingly, the ratio $I_{\rm H\beta}^{\rm Cer}/I_{\rm
H\beta}^{\rm nor}$ should be taken in a wider range, $I_{\rm
H\beta}^{\rm Cer}/I_{\rm H\beta}^{\rm nor}\leq 0.5$, rather than
$I_{\rm H\beta}^{\rm Cer}/I_{\rm H\beta}^{\rm nor}= 0$. Anyway, the
following conclusion remains beyond doubt: the ratio $I_{\rm
H\beta}^{\rm Cer}/I_{\rm H\beta}^{\rm nor}$ for the blend H$\beta$
line should be far from the `equipartition of intensity'. The small
values of $I_{\rm H\beta}^{\rm Cer}/I_{\rm H\beta}^{\rm nor}$ (in
range $I_{\rm H\beta}^{\rm Cer}/I_{\rm H\beta}^{\rm nor}\leq 0.5$)
imply that, comparing with Ly$\alpha$ and H$\alpha$ lines, the
Cerenkov component in the blend H$\beta$ line is no longer
important, rendering the `Cerenkov line redshift' difficult to be
detected (see Equation (\ref{expectedredshift}), and the first line
in Table \ref{tab01}).



\section{Conclusions and Discussions}
\label{sec:Conclusion}

Table \ref{tab01} shows that, the expected redshifts of the blended
hydrogen lines, given by model 1 or 2, are well fitted to the
observed values. Furthermore, comparing the observations and the
predictions of model 1 or model 2 (see the first, third and forth
lines in Table \ref{tab01}), we find that, both the observed and the
predicted redshifts have the same un-equality sequence, i.e., $
\Delta Z_{\rm Ly\alpha}^{\rm obs} > \Delta Z_{\rm H\alpha}^{\rm
obs}> \Delta Z_{\rm H\beta}^{\rm obs}$, as well as $ \Delta Z_{\rm
Ly\alpha}^{\rm exp} > \Delta Z_{\rm H\alpha}^{\rm exp}> \Delta
Z_{\rm H\beta}^{\rm exp}$. We stress that this un-equality sequence
originates from the intrinsic property of the Cerenkov hydrogen
lines, $ \Delta Z_{\rm Ly\alpha}^{\rm Cer} > \Delta Z_{\rm
H\alpha}^{\rm Cer}> \Delta Z_{\rm H\beta}^{\rm Cer}$, despite what
$T$-value is adopted (see Figure \ref{fig:1}). We claim that, the
good consistency, and the same un-equality sequence between
observations and predictions, are not coincident, instead, it is an
important evidence supporting the conclusion: the observed hydrogen
lines of quasars are blended by both the Cerenkov quasi-line and the
normal line; and the redshift-differences of different hydrogen
lines originate from the `Cerenkov line redshifts', existing in the
blended lines.

Another conclusion is that the average intensity ratios of the two
components in the blended hydrogen lines of $\sim$2,200 quasars are
not far from unit (approximate `equipartition of intensity'). We
particularly conclude that, for the strongest Ly$\alpha$ line,
$I_{\rm Ly\alpha}^{\rm Cer}/I_{\rm Ly\alpha}^{\rm nor}\approx 1$ is
more reliable because of the good fitting to the observed Ly$\alpha$
line (both in the blend model 1 and model 2, a more exact value
should be $I_{\rm Ly\alpha}^{\rm Cer}/I_{\rm Ly\alpha}^{\rm
nor}\approx 0.95$, given by blend model 2).
This could be a progress in study of the blend models for the broad
lines of quasars; the `equipartition of intensity' for Ly$\alpha$
line, the strongest one in all hydrogen lines, numerically confirms
the importance of the Cerenkov line-like radiation for the broad
lines of quasars, which can not be ignored any more.

The ratios $I_{\rm Ly\alpha}^{\rm Cer}/I_{\rm Ly\alpha}^{\rm
nor}=0.95$, $I_{\rm H\alpha}^{\rm Cer}/I_{\rm H\alpha}^{\rm
nor}=0.63$, and $I_{\rm H\beta}^{\rm Cer}/I_{\rm H\beta}^{\rm
nor}\leq 0.5$, adopted in the model 2, could be also helpful to
estimate the intensity-ratios of different blended hydrogen lines,
$I_{\rm Ly\alpha}^{\rm exp}/I_{\rm H\alpha}^{\rm exp}/I_{\rm
H\beta}^{\rm exp}$, where $I_{\rm Ly\alpha}^{\rm exp}\equiv I_{\rm
Ly\alpha}^{\rm Cer}+I_{\rm Ly\alpha}^{\rm nor}$ represents the total
intensity of the blend Ly$\alpha$ line, similarly for $I_{\rm
H\alpha}^{\rm exp}$, $I_{\rm H\beta}^{\rm exp}$. If the above
parameter-values $I_{\rm Ly\alpha}^{\rm Cer}/I_{\rm Ly\alpha}^{\rm
nor}=0.95$, $I_{\rm H\alpha}^{\rm Cer}/I_{\rm H\alpha}^{\rm
nor}=0.63$, and $I_{\rm H\beta}^{\rm Cer}/I_{\rm H\beta}^{\rm
nor}\leq 0.5$ (model 2) also give a good fitting to the observed
intensity-ratios $I_{\rm Ly\alpha}^{\rm obs}/I_{\rm H\alpha}^{\rm
obs}/I_{\rm H\beta}^{\rm obs}$,
the proposition of blended broad lines in quasars would be further
confirmed. This is worth to do in subsequent study.

The observed additional redshifts of the low-ionization lines,
relative to [O~III] 5007~{\AA} (e.g., Mg~II 2798~{\AA}, see Table 4
in \citet{Vanden-Berk01}), should be also arise from the `Cerenkov
line redshift', because the low-ionization lines also contain the
Cerenkov quasi-line component. We plan to fit the redshifts of some
low-ionization lines in the same way as for the hydrogen lines,
taking the intensity-ratios (e.g., $I_{\rm MgII}^{\rm Cer}/I_{\rm
MgII}^{\rm nor}$) as parameters.

In our blend model consideration, the `Cerenkov line shift' is the
unique source of the additional redshifts of hydrogen lines.
Therefore, the `blend models 1 and 2' are basically different from
the generally accepted `stratification models', in which the
additional redshifts all arise from the Doppler effect (the velocity
shift). The advantage of the `blend models' is obvious, which gives
the best fitting to the observed redshifts. We hope that the future
observations can give a further judgment on the two different kinds
of models.

As of now there remain some outstanding difficulties in the current
standard models to account for the observed broad hydrogen lines and
some low-ionization lines in quasars, such as the steep Balmer
decrement\citep[e.g.,][]{Osterbrock84, KK79, KK81, DU80, KG04}, the
different redshift values among some broad lines, the excess line
emission \citep[i.e., the energy-budget discrepancy of Fe II, Mg II,
He II lines in UV and optical wavebands,][]{Netzer85, Collin86,
MacAlpine03, Baldwin04, Joly08}. The quasi-line emission could
provide a solution for these long standing puzzles, for which the
energy is extracted from the kinetic energy of the fast electrons,
rather than from the continuum photons. We hope that the existence
of the component of this quasi-line emission in broad lines of
quasars can be tested by future observations.

The shape and redshift of emission lines in quasars reveal the
physical processes and activities of central engine around black
hole. Until now, Gaskell's results on the difference between low-
and high-ionization lines (represented by Mg~II (2800~\AA) and C~IV
(1549~\AA) respectively) have remained valid, and have yielded the
observational basis for a model that still provides a basic
interpretative sketch for the structure of broad line emitting
region \citep{Collin88}. In their model, the low-ionization
lines(LILs) and the hydrogen lines of quasars have a systemic
redshift relative to the high-ionization lines(HILs)
\citep{Gaskell82, Wilkes86, Corbin90, Corbin91, Espey+89,
Carswell+91, Sulentic+95, McIntosh+99}. They think that HIL blue
shifts are associated with radial motion in a flattened structure
(e.g., the accretion disk) mainly emitting LILs, also later
developments confirmed the earlier results \citep{CB96,
Marziani+96}. However, \citet{Sulentic+16} pointed out that not all
quasars are the same case, and that a first distinction between
radio-loud and radio-quiet quasars was necessary: blueshifts were
mostly present in radio quiet (RQ) quasars, but definitely rarer in
radio loud (RL) \citep[also~see,][]{Sulentic+00, Sulentic+07}. We
explain redshifts between the HILs and the LILs based on our primary
idea, which the HILs are mainly from the normal lines (i.e.,
bound-bound transition), whereas the LILs are blended by the
component of Cerenkov quasi-line, which has an intrinsic redshift
from line to line. Besides, this thought should be expected to be
confirmed that the Cerenkov quasi-line emission has much more
contribution to broad lines in RQ quasars because there may be much
denser gas ($N_{\rm H}\geq 10^{14}~{\rm cm}^{-3}$) required by
Cerenkov emission.

\acknowledgments We would like to thank the anonymous referee for
his/her careful reading of our manuscript which would help us to
improve it. DL acknowledges support by the National Science
Foundation of China (Grant Nos. 11078014 and 11125313), the key
laboratory grant from the Office of Science and Technology, Shanghai
Municipal Government (No. 11DZ2260700), the National Basic Research
Program of China (Grant Nos. 2009CB824904 and 2013CB837901), and the
Shanghai Science and Technology Commission (Program of Shanghai
Subject Chief Scientist; Grant Nos. 12XD1406200 and 11DZ2260700).







\end{document}